\documentclass[10pt,conference]{IEEEtran}
\IEEEoverridecommandlockouts
\renewcommand{\thetable}{\arabic{table}}%

\usepackage{hyperref}
\usepackage{url}
\usepackage{xurl}
\usepackage{graphicx}
\usepackage{textcomp}
\usepackage{xcolor}

\usepackage{caption}

\usepackage[ruled,vlined,linesnumbered]{algorithm2e}
\usepackage{pifont}
\usepackage{listings}
\usepackage{listings-rust}
\usepackage{xspace}
\usepackage{tabularray}
\usepackage[ff,mm,sets,adversary,advantage,asymptotics,probability,keys,primitives,operators,lambda,logic,notions]{cryptocode}
\usepackage[OT2,OT1]{fontenc}

\usepackage[flushleft]{threeparttable}
\usepackage{cuted}
\usepackage{boxedminipage}
\usepackage{amsfonts}
\usepackage{amsthm}
\usepackage{mathtools}
\usepackage[noend]{algpseudocode}
\usepackage[T1]{fontenc}
\usepackage{multirow}
\usepackage[table,xcdraw]{xcolor}
\usepackage[most]{tcolorbox}
\usepackage[normalem]{ulem}
\usepackage[]{minted}
\usepackage{enumitem}

\usetikzlibrary{patterns}

\graphicspath{{./figures/}}
\DeclareGraphicsExtensions{.pdf,.jpeg,.png}

\lstdefinestyle{mystyle}{
    commentstyle=\color{codegreen},
    keywordstyle=\color{magenta},
    numberstyle=\tiny\color{codegray},
    stringstyle=\color{codepurple},
    basicstyle=\ttfamily\footnotesize,
    breakatwhitespace=false,         
    breaklines=true,                 
    captionpos=b,                    
    numbers=none,                    
    showspaces=false,                
    showstringspaces=false,
    showtabs=false,                  
    tabsize=2
}

\def\Snospace~{\S{}}
\newcommand{\mystrut}{\strut}

\newcommand{\name}{\texttt{LineBreaker}\xspace}

\newcommand{\parhead}[1]{\vspace{5pt}\noindent {\textbf{#1}.}}
\newcommand{\mytilde}{\raise.17ex\hbox{$\scriptstyle\mathtt{\sim}$}}

\theoremstyle{definition}

\newcommand{\etc}{\emph{etc.}\xspace}
\newcommand{\ie}{\emph{i.e.,}\xspace}
\newcommand{\eg}{\emph{e.g.,}\xspace}

\newcommand{\hang}[1]{\textcolor{orange}{hang: #1}}

\newcommand{\ignore}[1]{}
\newcommand{\hongbo}[1]{}
\newcommand{\xun}[1]{}
\newcommand{\xing}[1]{}
\newcommand{\ronghua}[1]{}
\newcommand{\yifan}[1]{}

\newcommand{\luyiComment}[1]{}
\newcommand{\LX}[1]{}

\newcommand{\sotalm}{GPT-4\xspace}

\author{
    \IEEEauthorblockN{
        Hongbo Chen\IEEEauthorrefmark{1}\IEEEauthorrefmark{2}, 
        Yifan Zhang\IEEEauthorrefmark{1}\IEEEauthorrefmark{3}\IEEEauthorrefmark{2}\IEEEauthorrefmark{8}, 
        Xing Han\IEEEauthorrefmark{1}\IEEEauthorrefmark{4}, 
        Tianhao Mao\IEEEauthorrefmark{2}, 
        Huanyao Rong \IEEEauthorrefmark{2}, \\
        Yuheng Zhang \IEEEauthorrefmark{6}, 
        XiaoFeng Wang \IEEEauthorrefmark{7}, 
        Luyi Xing \IEEEauthorrefmark{2} \IEEEauthorrefmark{5}, 
        Xun Chen \IEEEauthorrefmark{9} \IEEEauthorrefmark{8}, 
        Hang Zhang \IEEEauthorrefmark{2}}
    \IEEEauthorblockA{
        \IEEEauthorrefmark{2}Indiana University Bloomington, USA 
        \IEEEauthorrefmark{3}San Diego State University, USA \\
        \IEEEauthorrefmark{4}The Hong Kong University of Science and Technology, China 
        \IEEEauthorrefmark{5}University of Illinois Urbana-Champaign, USA \\
        \IEEEauthorrefmark{6}Tsinghua University, China
        \IEEEauthorrefmark{7}Nanyang Technological University, Singapore
        \IEEEauthorrefmark{9}Independent researcher \\
        \{hc50, tianmao, ronghua, hz64\}@iu.edu, yzhang24@sdsu.edu, xhanbg@cse.ust.hk, zhangyuh25@mails.tsinghua.edu.cn, \\
        xiaofeng.wang@ntu.edu.sg, lxing2@illinois.edu, xunchen@outlook.com
    }
    \thanks{\IEEEauthorrefmark{1}Equally contributed to this work.}
    \thanks{\IEEEauthorrefmark{8} These two authors were in Samsung Research America.}
    }

\def\BibTeX{{\rm B\kern-.05em{\sc i\kern-.025em b}\kern-.08em
    T\kern-.1667em\lower.7ex\hbox{E}\kern-.125emX}}
\begin{document}

\title{\name: Finding Token-Inconsistency Bugs with Large Language Models}

\maketitle

\begin{abstract}
Token-inconsistency bugs (TIBs) involve the misuse of syntactically valid yet incorrect code tokens, such as misused variables and erroneous function invocations, which can often lead to software bugs.
Unlike simple syntactic bugs, TIBs occur at the semantic level and are subtle - sometimes they remain undetected for years.
Traditional detection methods, such as static analysis and dynamic testing, often struggle with TIBs due to their versatile and context-dependent nature.
However, advancements in large language models (LLMs) like GPT-4 present new opportunities for automating TIB detection by leveraging these models' semantic understanding capabilities.

This paper reports the first systematic measurement of LLMs' capabilities in detecting TIBs,
revealing that while GPT-4 shows promise, it exhibits limitations in precision and scalability. 
Specifically, its detection capability is undermined by the model's tendency to focus on the code snippets that do not contain TIBs; its scalability concern arises from GPT-4's high cost and the massive amount of code requiring inspection.
To address these challenges, we introduce \name, a novel and cascaded TIB detection system. \name leverages smaller, code-specific, and highly efficient language models to filter out large numbers of code snippets unlikely to contain TIBs, thereby significantly enhancing the system's performance in terms of precision, recall, and scalability. 
We evaluated \name on 154 Python and C GitHub repositories, each with over 1,000 stars, uncovering 123 new flaws, 45\% of which could be exploited to disrupt program functionalities.
Out of our 69 submitted fixes, 41 have already been confirmed or merged. 

\end{abstract}


\definecolor{myblue}{RGB}{70,130,180}
\definecolor{mydeepblue}{RGB}{65,105,225}
\definecolor{myviolet}{RGB}{97,0,138}
\definecolor{myburgundy}{RGB}{110,10,30}
\definecolor{mygreen}{RGB}{0,105,148}
\definecolor{mygrey}{RGB}{180, 180, 200}
\definecolor{idealfun}{RGB}{165,42,42}
\definecolor{check}{RGB}{11,141,10}
\definecolor{auburn}{rgb}{0.43, 0.21, 0.1}
\definecolor{codegreen}{rgb}{0,0.6,0}
\definecolor{codegray}{rgb}{0.5,0.5,0.5}
\definecolor{codepurple}{rgb}{0.58,0,0.82}
\newcommand{\bnf}[1]{\langle {\tt {#1}} \rangle}
\newcommand{\addr}[1]{\ifmmode{\tt 0x#1}\else ${\tt 0x#1}$\fi}
\newcommand{\choice}{\; \mid \;}

\def\Snospace~{\S{}}
\renewcommand{\sectionautorefname}{\Snospace}
\renewcommand{\subsectionautorefname}{\Snospace}
\renewcommand{\subsubsectionautorefname}{\Snospace}
\renewcommand{\appendixautorefname}{\Snospace}

\section{Introduction}
\label{sec:intro}

Semantic bugs can cause significant damage in real-world systems, leading to unexpected behaviors, security vulnerabilities, and operational failures. 
Yet, these bugs are notoriously difficult to detect because they often require a deep understanding of program logic and context.
In this work, we focus on the automatic and scalable detection of Token-Inconsistency Bugs (TIBs)\footnote{To differentiate between lexical tokens in programming languages (e.g., variable names) and basic text units in natural language processing (e.g., subwords), we use the terms "code token" and "text token" respectively.
}, a subclass of semantic bugs 
which is challenging for existing approaches.

\noindent\textbf{Token-Inconsistency Bugs}. 
TIBs are introduced when syntactically valid code tokens — such as variables, calls, and operators — are placed incorrectly, leading to subtle yet impactful flaws.
For example, \autoref{code:method-misuse} demonstrates a TIB detected by our tool in WolfSSL - a security-sensitive project with over 2.3K GitHub stars,
where a redundant \texttt{sizeof()} call leads to incorrect buffer size and memory errors.
In \autoref{fig:example1}, \texttt{quote()} mistakenly takes the argument \texttt{query} rather than \texttt{params}, disrupting normal program logic.
These syntactically valid but semantically incorrect bugs are frequently introduced by developers~\cite{tan2014bug},
leading to several security consequences such as control/data flow integrity violations, denial of service, \etc~\cite{khoshnoud2022bugs,hwang2020gap},
evidenced by CVEs assigned to TIBs (\eg CVE-2022-30780, CVE-2022-23574, and CVE-2020-8889).

% ...

\begin{figure}[ht] 
    \begin{lstlisting}[
        language=C,
        basicstyle=\fontsize{6.68pt}{8.68pt}\selectfont\ttfamily, % Set font size and family
        escapeinside={|}{|}, % The key part: use | as escape characters
        caption={An example TIB: \texttt{sizeof(xSize)} in red is buggy; the correct code \texttt{xSize} is in green.},
        label={code:method-misuse},
        captionpos=b, % Position caption at the bottom
        breaklines=true % Allow lines to wrap
    ]
socklen_t xSize = sizeof(struct freertos_sockaddr);
xSocket_t xClientSocket = NULL;
struct freertos_sockaddr xRemoteAddress;
...
FreeRTOS_bind(xClientSocket, &xRemoteAddress,
    |\colorbox{red!30}{\mystrut\sout{sizeof(xSize)}}||\colorbox{green!50}{\mystrut{xSize}}|);
    \end{lstlisting}
    \vspace{-4mm} 
\end{figure}

\begin{figure}[ht]
\footnotesize
\begin{tcolorbox}[colback=gray!10!white, 
                  colframe=black, 
                  fonttitle=\footnotesize,
                  fontupper=\fontsize{6.68pt}{8.68pt}\selectfont,
                  fontlower=\fontsize{6.68pt}{8.68pt}\selectfont,
                  arc=2mm, 
                  boxsep=5pt, 
                  left=0pt, 
                  right=0pt, 
                  top=0pt,
                  bottom=0pt,
                  boxrule=0.5pt 
                  ]
\textbf{Prompt}: Find bugs in this Python code snippet.
\begin{lstlisting}[
    language=Python,
    basicstyle=\fontsize{6.68pt}{8.68pt}\selectfont\ttfamily,
    escapeinside={|}{|}
]
def quote_url(url_text):
    (scheme, netloc, path, params, query, fragment) =
        six.moves.urllib.parse.urlparse(url_text)
    # netloc_quoted = quote(netloc)
    path_quoted = quote(path)
    params_quoted = quote(|\colorbox{red!30}{\mystrut\sout{query}}||\colorbox{green!50}{\mystrut{params}}|)
    query_quoted = quote_plus(query)
    fragment_quoted = quote(fragment)
    url_quoted = six.moves.urllib.parse.urlunparse(...)
    return url_quoted
\end{lstlisting}
\tcblower
\textbf{Output by ChatGPT-4}:
The code snippet you provided has a couple of issues that could be causing problems:

1. \textbf{Incorrect usage of the \texttt{quote} function for \texttt{params}}:
In the code, \texttt{params\_quoted = quote(query)}. This line mistakenly passes the \texttt{query} to the \texttt{quote} function instead of \texttt{params}. It should be \texttt{params\_quoted = quote(params)}. \\
2. \textbf{Potential Misuse of \texttt{quote} vs. \texttt{quote\_plus}}:
Using \texttt{quote\_plus} for the query is appropriate, but if any other component requires handling spaces as \texttt{+}, then \texttt{quote\_plus} should be used accordingly. Conversely, if spaces should be encoded as \texttt{\%20} in the query, then \texttt{quote} should be used instead of \texttt{quote\_plus} \\
\textit{Another false alarm and a corrected code snippet are truncated.}
\end{tcolorbox}
\caption{An Example of ChatGPT detecting an existing bug. The variable \texttt{query} in the red rectangular is buggy.}
\label{fig:example1}
\end{figure}



\noindent\textbf{Difficulties in TIB Detection.}
Despite their security consequences, TIBs often elude conventional detection methods.
As in \autoref{fig:example1}, although a human reviewer might identify the misplaced pair (\texttt{\textbf{query}} and \texttt{\textbf{params}\_quoted}) from the program's context (\texttt{path} vs. \texttt{path\_quoted} and \texttt{query} vs. \texttt{query\_quoted}), the buggy code lacks clear-cut patterns as observed in vulnerabilities like use-after-free.
Static analysis struggles to capture such context-dependent bugs without explicit rules, and dynamic testing (\eg fuzzing) may fail due to limited code coverage or subtle bug symptoms (\eg logic errors may not cause program crashes that are captured by fuzzers).
As a result, TIBs can remain undetected for years.
The TIB shown in \autoref{fig:example1} had persisted in a popular GitHub repository for seven years~\cite{rabbitvcs_pull_request},
and the bug in \autoref{code:method-misuse} went unnoticed for four years.


\vspace{2pt}\noindent\textbf{LLM-based TIB Detection \& Challenges.}
Human developers may identify TIBs by leveraging their capability to predict the ``correct'' code token based on the program's context, such as the correlations between argument names and function names in \autoref{fig:example1} and between variable and function semantics in \autoref{code:method-misuse}.
Recent advances in large language models (LLMs) offer a similar predictive capacity: given a partial code, the LLM can predict the most likely subsequent tokens.
Any inconsistency between the actual code and the prediction result could indicate a potential TIB.
Moreover, unlike previous ML-based semantic bug detectors that need training on datasets tailored to specific programming languages~\cite{allamanis2018learning, vasic2018neural, pradel2018deepbugs}, LLMs are pre-trained on massive corpora, providing them with \textit{generality} across different languages and \textit{flexibility}, often requiring little to no additional training.
For example, on our constructed benchmark dataset, which includes 100 synthesized TIBs in 5.7K lines of Python code, GPT-4 can capture up to 72\% of them with proper prompts (\autoref{sec:measurement:gpt}).
Despite promising preliminary results,
LLMs are not ready for comprehensive, highly scalable, and reasonably accurate TIB detection.
Current models typically restrict the input/output token length, making it difficult to incorporate the entire code base into them.
For example, Pytorch has roughly 1.4 million lines of code, which can overwhelm GPT-4o's 128k context length.
However, slicing large codebases and iteratively prompting capable off-the-shelf LLMs can be prohibitively expensive (\eg the API cost increases quadratically with code size under an LLM-based detector, FLAG~\cite{flag}). 


Furthermore, LLMs' answers can be error-prone,
as pointed by OpenAI~\cite{gpt4} and corroborated by our measurements,
leading to many false reports in TIB detection; half of the functions sampled from popular GitHub repositories are marked ``buggy'' by GPT-4. 
Often, these reports occur because the model is \textit{distracted} by code contexts unrelated to the targeted flaws~\cite{sun2024llm4vuln}.
For example, in \autoref{fig:example1}, GPT-4 flags \texttt{quote\_plus} incorrectly, probably due to it being distracted by \texttt{plus}.
\vspace{2pt}\noindent\textbf{Our Solution}.
To address those challenges, we introduce \textit{lightweight} models for TIB detection and \emph{cascade} models of different sizes, capabilities, and efficiency in a unified analytical model to seek a balance between accuracy, cost, and effectiveness through optimal selection (\autoref{sec:design:cascaded}). 
We also improve LLMs' performance on TIB detection using meticulously designed prompt structures (\autoref{sec:design:hl}) and token generation algorithms (\autoref{sec:design:controlled}), as motivated by the measurement.


Exploring the design space through this analytical model, we propose \name, a TIB detection pipeline (\autoref{sec:design}).
It first performs static analysis, enumerating code tokens that may involve TIBs.
Then, \name utilizes lightweight language models to filter potential TIBs locally.
We compare predicted tokens with the original tokens, and
any difference indicates a potential TIB worth further inspection.
Applying such a filter several times efficiently excludes most TIB-free positions, boosting the TIB density and detection precision.
Finally, the remaining suspicious TIB sites are uploaded to a SOTA model for in-depth analysis.
This cascaded design significantly reduces costs, since smaller, faster, and cheaper models filter out most faults. 
It also improves the precision of TIB detection, as the final SOTA model can thoroughly analyze a much smaller yet more relevant set of potential buggy locations, mitigating the distraction problem.
\vspace{2pt}\noindent\textbf{Findings}. We extensively evaluated \name on large-scale real-world datasets.
\name successfully discovered 123 new TIBs in popular C and Python repositories on GitHub with more than 1K stars.
We submitted 69 pull requests, and developers have acknowledged 41 of them.
\name also achieves an improved precision (up to 36.36\%) compared to an existing technique for detecting semantic bugs in real-world code (only 12.0\%~\cite{ahmadi2021finding}).
It also maintains practical monetary cost, spending only \$92.
Furthermore, the explanation provided by the LLM pipeline enables a human expert to conveniently identify the true bugs in the filtered code snippets. 
We responsibly reported all bugs and vulnerabilities to the relevant developers.



\vspace{2pt}\noindent\textbf{Contributions}. We summarize our contributions as follows.

\vspace{2pt}\noindent$\bullet$~\textbf{Study} (\autoref{sec:measurement}).
    We report the first comprehensive measurement study on the capabilities of various LLMs in TIB detection, shedding light on their strengths and limitations. 
    
\vspace{2pt}\noindent$\bullet$~\textbf{System} (\autoref{sec:design}).
    Based upon our measurement study and analytical model, we develop \name, a cascaded pipeline integrated with new mechanisms to enhance LLMs' performance in TIB detection.
    \name is both practical and efficient for various languages.
    The \name is publicly available.\footnote{https://github.com/Gao-Chuan/LineBreaker}
    
\vspace{2pt}\noindent$\bullet$~\textbf{Findings} (\autoref{sec:evaluation}).
    We use \name to perform the first TIB detection in real-world code at scale, detecting 90 and 33 previously unknown TIBs, respectively, from popular Python and C repositories.
    45\% of them have various security implications.  
    We have submitted 69 fixes, of which 41 have been confirmed or merged.
    
\vspace{2pt}\noindent$\bullet$~\textbf{Dataset}.
    We curated a comprehensive dataset of TIBs from historical bug fixes, new bugs discovered by \name, and our synthesized ones.
    We will release it for future research.
\section{Measurement Study}
\label{sec:measurement}

To optimally leverage LLMs for TIB detection, it is essential to understand their performance in this specific task.
In this section, we comprehensively evaluate the performance of various language models, encompassing the SOTA model and smaller open-source models designed for coding tasks.

\subsection{Language-Model based TIB Detection}
\label{sec:codelm}

\parhead{Instruct Large Language Models}
Represented by OpenAI's GPT-4 model family~\cite{gpt4}, instruct models are trained on vast, diverse corpora. 
This allows them to process tasks effectively in various domains, including software security.
However, they are often proprietary (\eg GPT-4) and expensive to use at scale, restricting their adaptability and making experimental modifications infeasible.
Furthermore, these models are also prone to producing inaccurate or fabricated responses due to distraction and well-documented ``hallucination'', leading to a high false rate in TIB detection.
The false positive case in \autoref{fig:example1} exemplifies this concern.

\parhead{Code-specific Language Models}
Recent efforts have produced models specialized for code-related tasks, such as code generation and completion. 
These models share a similar transformer-based architecture that generates token probabilities based on context.
CodeBERT is a bidirectional encoder model pre-trained on masked token infilling tasks~\cite{codebert}, enabling it to predict masked tokens in the middle of a snippet.
However, its context length is strictly limited to 512 tokens, making it impractical for real-world applications.
Recent models adopt a decoder-only transformer design, generating tokens sequentially.
Despite their unidirectional nature, some are trained on fill-in-the-middle (FIM) tasks, enabling code infilling like CodeBERT.
While not as powerful as GPT-4 in some respects, these open-source code-specific models offer broader accessibility and opportunities for customization.

\begin{figure}
\center
\includegraphics[width=0.48\textwidth]{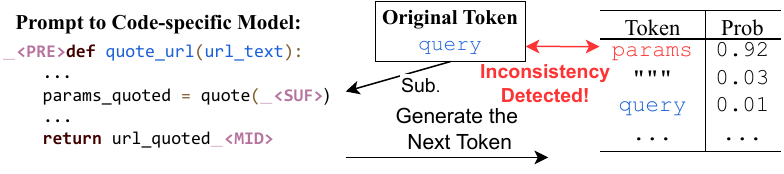}
\caption{TIB detection using code-specific LLM.}
\label{fig:tokengen}
\end{figure}

\noindent\textit{Example.}
Code-infilling can naturally and novelly support TIB detection by comparing the original tokens against a model’s predicted tokens.
If the predicted tokens differ, this indicates a potential bug.
For instance, when feeding the buggy code snippet in \autoref{fig:example1} into a code-specific model supporting FIM as in  \autoref{fig:tokengen},
the model recommends \texttt{params} as a more contextually appropriate token, strongly suggesting a TIB.

\subsection{Dataset Preparation}
\label{sec:measurement:dataset}



Since no dedicated datasets for TIBs exist, we constructed bipartite datasets to evaluate key metrics, including recall and specificity.
To mitigate memorization and contamination concerns, we sampled code snippets from popular (>1k stars) Python repositories on GitHub committed after the LLMs' knowledge cutoff date.
Functions were grouped by length and proportionally sampled to maintain a comparable context length distribution. 
The resulting dataset, $D$, contains 1,000 manually verified functions, all of which are free of TIBs.



We also implement a process akin to prior code token mutation methodologies~\cite{richter2022learning, patra2021semantic} to introduce synthesized TIBs on the dataset $D$, obtaining a synthesized dataset $D'$.
$D'$ comprises artificial yet plausible TIBs suitable for, \eg recall evaluation.
This process is initiated by enumerating all permissible code token candidates that could feasibly replace another token without causing syntax errors.
Permissible candidates span over four types of code tokens: variable uses, function calls, operators (\eg $>=$ and \texttt{or}), and literals (\eg integer).
We then select a substituted TIB token with slight semantic divergence based on embeddings from MPNet~\cite{song2020mpnet}, reflecting common confusion between tokens with similar semantics.
Function snippet in $D'$ has a single TIB.
Three security experts independently validated the datasets and then convened to discuss and resolve all identified inaccuracies.


\subsection{GPT-4's Performance on TIB Detection}
\label{sec:measurement:gpt}

To assess the TIB detection capabilities of general-purpose LLMs, we selected GPT-4 models
for their superior performance~\cite{sun2024llm4vuln, steenhoek2024comprehensive, wei2024demystifying} and API's support for JSON output.



\begin{figure}[ht]
\begin{tcolorbox}[colback=gray!10!white, 
                  colframe=black, 
                  fonttitle=\fontsize{6.68pt}{8.68pt}\selectfont,
                  fontupper=\fontsize{6.68pt}{8.68pt}\selectfont,
                  fontlower=\fontsize{6.68pt}{8.68pt}\selectfont,
                  arc=2mm, 
                  boxsep=5pt, 
                  left=0pt, 
                  right=0pt, 
                  top=-2pt,
                  bottom=-2pt,
                  boxrule=0.5pt 
                  ]
\footnotesize
\textbf{System}:
You're a {language} expert. Your job is to inspect if the code contains any semantic bugs. Semantic bugs are a type of bug that occurs when the code is syntactically correct but does not behave as intended or produces incorrect results. These bugs arise from mismatches between the programmer's intended logic and the actual implementation in the code, where incorrect variable/method name usage or assignment can lead to bugs or vulnerabilities in the program.

Implement your work by the following steps:
\begin{enumerate}[label=\arabic*., leftmargin=2em]
\item Check the given code line by line.
\item For each line, extract the variables, calls, operators, and literals in this line.
\item Check whether the extracted variables, calls, operators, and literals are correctly used in the statements.
\item If there is no wrong usage in any line, return the code is correct; otherwise, record each line of the bug and the reason for the bug.
\end{enumerate}

Assume the code is syntactically correct and input parameters to the functions are well-formed and valid. Focus solely on detecting semantic bugs, and ignore other problems. Return your response in JSON format.

\tcblower
\textbf{User Prompts} \\
\textbf{Round 1 Properties}: Mandatory: {\texttt{\{code\_line, explanation\}}} \\
\textbf{Round 1 User Prompt}: \texttt{\{code for detection\}}
Output exact lines of semantic bugs and concise explanations. \\
\textbf{Round 2 Properties}: Mandatory: {\texttt{\{code\_line, explanation\}}; \\
Selective: \texttt{\{fixed\_line}, token\_level, category\}} \\
\textbf{Round 2 User Prompt}: Inspect these bugs, excluding 1. incorrect or unlikely bugs; 2. non-semantic bugs. Check the left bugs only break the intended functionality or lead to vulnerabilities. Finish the following tasks: 1. Answer if the bug is related to a single or a few tokens. 2. Using the code and previous contexts, classify the bugs into these categories: Security Vulnerability, Logic Bug, Enhancement, Unexpected Behavior, Symbol Not Defined, Module Not Imported, Bad Smell, Not a Bug, or Others. Note that the snippet is from popular repositories and runs, so correct symbols not defined in the snippet might be defined at other places, and you can regard it as Symbol not Defined. If you assign Others category, explicitly name the category.
\end{tcolorbox}
\caption{Example: a two-round prompt template (1/2FTCa).}
\label{fig:prompt-12FMCa}
\end{figure} 

\parhead{Prompt Design}
In-context examples do not significantly improve GPT-4's accuracy in identifying vulnerabilities~\cite{steenhoek2024comprehensive}.
Therefore, given the diverse nature of TIB patterns, we opted for zero-shot prompts to avoid the influence of potentially irrelevant examples~\cite{shi2023large}. 
Specifically, we utilized various templates and conducted multiple rounds of interactive prompts.
These templates incorporate various prompt engineering techniques, including Chain-of-Thought (CoT)~\cite{wei2022chain} and cross-examination~\cite{cohen2023lm}. Drawing on the prompt engineering methods introduced in~\cite{White2023APP}, we assigned the role of a programmer to the LLM and provided it with an output pattern.
Additionally, following the prompt taxonomy and templates proposed in~\cite{karmaker-santu-feng-2023-teler}, we outlined the workflow steps to offer detailed instructions to the LLM on effectively processing the given prompt.\hongbo{metaprompt}


%
%
%
%

\begin{table*}[t!]
    \centering
    \caption{Analytical properties for TIB detection and inspection in JSON-formatted tool calling.}
    \fontsize{6.68pt}{8.68pt}\selectfont
\begin{tabular}{p{0.12\linewidth}|p{0.04\linewidth}|p{0.06\linewidth}|p{0.3\linewidth}|p{0.2\linewidth}|p{0.09\linewidth}}
\hline
\textbf{Property}                 & \textbf{Abbr.} & \textbf{Type}   & \textbf{Description}                                                                      & \textbf{Purpose}      & \textbf{Prompt Tec.}                                        \\ \hline
\rowcolor{gray!20}\texttt{\textbf{co}de\_line}      & /     & Mandatory         & the exact line of code with semantic bug                                                  & locating the bug                  & /                          \\ 
\texttt{\textbf{e}xplanation}     & /     & Mandatory     & a concise explanation of this bug                                                         & reasoning about of the bug        & /                           \\ 
\rowcolor{gray!20} \texttt{\textbf{t}oken\_level} & T     & Selective     & the bug appears is related to single/few tokens             & filtering out non-TIB bugs & Cross-exam.  \\ 
\texttt{\textbf{f}ixed\_line}     & F    & Selective          & the line with the bug fixed                                                               & CoT for \texttt{token\_level} and fixing        & CoT    \\ 
\rowcolor{gray!20} \texttt{\textbf{ca}tegory}        & Ca     & Selective        & the category of this bug  & filtering out non-semantic bugs            & Cross-exam.                 \\
\texttt{\textbf{p}riority}        & P      & Selective        & the priority level of this bug: high/medium/low                                           & filtering out less interesting bugs            & Cross-exam.             \\ \hline
\end{tabular}
\label{tab:property} 
\\
\end{table*}

We illustrate an example of a template configuration in \autoref{fig:prompt-12FMCa}, which represents a simplified version of the one ultimately adopted by \name. 
The system prompt instructs the model to assume the role of a programming language expert (\ie persona adoption~\cite{openai:prompt-engineering}).
In the initial round, the model is requested to identify potential buggy code lines and provide explanations. 
The second round further cross-examines the identified potential TIBs to minimize false positives.
Additionally, we incorporate three selectable properties in this round: self-evaluated bug categorization, fix proposal, and token-level predicate.
The details of these properties are summarized in \autoref{tab:property}.
Mandatory properties are adopted to identify bugs, while selective properties are employed for filtering.
Multi-round prompting enhances effectiveness; for instance, allowing GPT-4 to reflect on its prior results, facilitating the exclusion of potential false alarms (\ie cross-examination). 
Depending on the properties required in each round, prompts for querying these properties can be automatically generated under various template setups.
If the property in \autoref{tab:property} is selected in a round, the digits represent the number of the round, and the alphabetic abbreviations represent chosen properties. 
For instance, the template in \autoref{fig:prompt-12FMCa} is denoted as 1/2FTCa.
Please refer to our supplementary material for other prompt templates.

\begin{table*}[]
\centering
\caption{GPT-4o TIB detection on different prompt templates.}
\fontsize{6.68pt}{8.68pt}\selectfont
\begin{tabular}{l|clclc|ccllc|c}
\hline
\multirow{2}{*}{Template} & \multicolumn{5}{c|}{Synthesized $\subset D'$}                                               & \multicolumn{5}{c|}{Bug-free $\subset D$}                                                   & \multirow{2}{*}{Cost} \\
                          & $TP_{L}$ & $FP_{L}$                 & $TP_{F}$    & $FP_{F}$                & Rec.          & $FP_{L}$ & E. $FP_{L}$              & $FP_{F}$    & E. $FP_{F}$             & Spe.          &                       \\ \hline
1                         & 81       & \multicolumn{1}{l|}{203} & \textbf{72} & \multicolumn{1}{l|}{28} & \textbf{72.0} & 404      & \multicolumn{1}{c|}{0}   & 29          & \multicolumn{1}{l|}{0}  & \textbf{0.0}  & \$0.83                \\
1FT                       & 75       & \multicolumn{1}{l|}{110} & 71          & \multicolumn{1}{l|}{24} & 71.0          & 229      & \multicolumn{1}{c|}{67}  & 14          & \multicolumn{1}{l|}{7}  & 22.6          & \$0.88                \\
1/2FT                     & 79       & \multicolumn{1}{l|}{79}  & 67          & \multicolumn{1}{l|}{33} & 67.0          & 298      & \multicolumn{1}{c|}{89}  & 24          & \multicolumn{1}{l|}{7}  & 23.0          & \$2.11                \\
1/2FT/3P                  & 57       & \multicolumn{1}{l|}{99}  & 55          & \multicolumn{1}{l|}{36} & 55.0          & 211      & \multicolumn{1}{c|}{73}  & 20 & \multicolumn{1}{l|}{10} & \textbf{25.7} & \$3.37                \\
1/2FT/3Ca                 & 72       & \multicolumn{1}{l|}{129} & 66          & \multicolumn{1}{l|}{31} & 66.0          & 245      & \multicolumn{1}{c|}{43}  & 19          & \multicolumn{1}{l|}{10} & 14.9          & \$3.43                \\
1/2FTCa                   & 78       & \multicolumn{1}{l|}{126} & \textbf{72} & \multicolumn{1}{l|}{26} & \textbf{72.0} & 294      & \multicolumn{1}{c|}{92}  & \textbf{23}          & \multicolumn{1}{l|}{7} & \textbf{23.8} & \$2.14                \\ \hline
1/2FTCa w/ HL             & 92       & \multicolumn{1}{l|}{93}  & \textbf{84} & \multicolumn{1}{l|}{12} & \textbf{84.0} & 174      & \multicolumn{1}{c|}{138} & \textbf{56} & \multicolumn{1}{l|}{34} & \textbf{44.2} & \$2.07                \\ \hline
\end{tabular}
\label{tab:gpt4-templates}

Rec. refers to \textit{recall} at the function-level and Spe. refers to filtering capability (i.e., \textit{specificity}); $X_{L}$ and $X_{F}$ represent the numbers counted \\ at the \textit{line}- and \textit{function}-level; $FP$ and E. $FP$ represent the numbers of incorrect bugs reported and \textit{excluded}.
\end{table*}

\begin{table*}[]
\centering
\caption{TIB detection performance of GPT-4 model family.}
\fontsize{6.68pt}{8.68pt}\selectfont
\begin{tabular}{l|clclc|ccllc|c}
\hline
\multirow{2}{*}{Model} & \multicolumn{5}{c|}{Synthesized $\subset D'$}                                               & \multicolumn{5}{c|}{Bug-free $\subset D$}                                                  & \multirow{2}{*}{Cost} \\
                       & $TP_{L}$ & $FP_{L}$                 & $TP_{F}$    & $FP_{F}$                & Rec.          & $FP_{L}$ & E. $FP_{L}$             & $FP_{F}$    & E. $FP_{F}$             & Spe.          &                       \\ \hline
GPT-4o                 & 78       & \multicolumn{1}{l|}{126} & \textbf{72} & \multicolumn{1}{l|}{26} & \textbf{72.0} & 294      & \multicolumn{1}{c|}{92} & 23          & \multicolumn{1}{l|}{7} & 23.8          & \$2.14                \\
GPT-4 Turbo            & 86       & \multicolumn{1}{l|}{129} & \textbf{72} & \multicolumn{1}{l|}{24} & \textbf{72.0} & 256      & \multicolumn{1}{c|}{87} & \textbf{19} & \multicolumn{1}{l|}{10} & \textbf{25.4} & \$6.82                \\
GPT-4o Mini            & 86       & \multicolumn{1}{l|}{242} & 64          & \multicolumn{1}{l|}{33} & 64.0          & 388      & \multicolumn{1}{c|}{92} & 29          & \multicolumn{1}{l|}{4}  & 19.2          & \$0.19                \\ \hline
\end{tabular}
\label{tab:gpt-models}
\end{table*}

\parhead{Measurement Results}
Our objective is to assess the performance of various prompt templates regarding recall, specificity, and monetary cost. 
While each function in $D'$ contains only one known TIB, the LLM may identify multiple TIBs within the queried function. 
We adopt two statistical approaches: 1) counting the number of lines flagged with TIBs. Flagged lines are regarded as positive, as the model is prompted to mark buggy lines; 2) counting the number of entire functions, as the security engineer usually audits the function as a whole. The metrics under these two approaches are subscripted by $_L$ and $_F$, respectively.
To reduce the high API costs, we evaluate all templates on subsets of $D'$ and $D$, each containing 100 cases.
%

\autoref{tab:gpt4-templates} presents the measurement results of prompt templates. 
We observe a high false positive rate when using a single-round template with no selective properties, and GPT-4 is directly asked to identify TIBs.
This is primarily due to distractions, as confirmed by selective manual inspection.

First, we unveil the effectiveness of selective properties (in \autoref{tab:property}). 
Our findings indicate that the \texttt{token\_level} property is particularly effective at filtering out non-TIB instances, which are not bugs associated with a few code tokens. 
This filtering improves line-level specificity by up to 22.6\% at the cost of a 1\% reduction in recall.
Additionally, excluding candidates whose \texttt{category} is not in \{Logic Bug, Security Vulnerability, Bad Smell\} further enhances specificity. 
However, filtering based on \texttt{priority $\ne$ high}, while reducing false positives, leads to missed bugs. 
This trade-off may be attributed to GPT-4's limited understanding of bug severity.

Next, we investigate the impact of iterative rounds. 
Introducing second and third rounds with varied properties yields mixed results, with cross-examination occasionally reducing false positives. 
However, adding a third prompt round proves less advantageous regarding specificity, recall, and monetary cost, as additional rounds increase token usage. 
This can be attributed to superfluous cross-examination, where identified bugs are unnecessarily filtered out. 
Distraction may also be a potential root cause, as irrelevant information from previous rounds can misguide the LLM. 
As a result, two-round prompting emerges as the optimal approach.

Furthermore, we evaluate the best overall prompt template, 1/2FTCa, on three models in the GPT-4 family, as shown in \autoref{tab:gpt-models}. 
Interestingly, GPT-4 Turbo slightly outperforms the newer GPT-4o on specificity with triple the cost.

\ignore{
On the other hand, we examine the effect of iterative prompting.
Introducing a second round improves the recall by approximately 4\% from 1FT to 1/2FT and specificity by 38\%,
highlighting that cross-examination can effectively reduce false positives.
Notably, adding a third prompt round proves less beneficial regarding specificity, recall, and monetary cost (\ie extra rounds result in more tokens).
This can be attributed to superfluous cross-examination, where bugs are identified but filtered out.
The root underlying cause could be distraction~\cite{sun2024llm4vuln}, wherein irrelevant information from the first and second rounds misleads the LLM.
Therefore, two-round prompting appears to be the optimal approach for TIB detection.
}


\parhead{Conclusion}
We arrive at two primary conclusions:

\noindent\textit{(1) Template Selection.}
The template 1/2FTCa in \autoref{fig:prompt-12FMCa} shows anticipated overall performance.
\name adopted it with suspicious lines highlighted (\autoref{sec:design:hl}).


\noindent\textit{(2) Limitations of GPT-4 in TIB Detection.}
Our study reveals that GPT-4 demonstrates relatively high recall for TIB detection. 
However, its current performance remains insufficient for practical use.
Even with the most effective ``1/2FTCa'' template, a significant portion of TIBs is still missed. 
Moreover, as TIBs in real-world code are relatively rare, relying solely on GPT-4 for TIB detection would likely result in a high false positive rate due to its limited specificity. 
As shown by the specificity evaluated on $D$, many false positive reports persist.
Our investigation into some cases reveals that GPT-4 is overly cautious when evaluating software security, resulting in the generation of false alarms.
Consequently, more targeted guidance is necessary to enhance the model's performance.



\subsection{Code-Specific Models on TIB Detection}
\label{sec:measurement:infilling}


While general-purpose LLMs can also perform code infilling, their usage cost can quickly become prohibitive for large codebases.
For instance, GPT-4 costs approximately  \$2.1 to infill every code token for a function of 40 lines of code,
which increases quadratically with code size.
Therefore, we focus on smaller open-source code language models in this evaluation.
The example in \autoref{sec:codelm} indicates that the code-infilling capability of LLMs can also facilitate TIB detection.
TIB detection could leverage infilling because the probabilities of correct tokens are significantly higher than those of buggy tokens.
We thus evaluate the performance of various code-specific LLMs on TIB detection.

\parhead{Hardware and Parameters}
We conduct the experiments on a Linux server with two AMD EPYC 9124 CPUs and one NVIDIA H100 GPU. The temperature is set to 0.

\parhead{Infilling Task Description}
Code infilling is a foundational capability of these code-specific models, enabling numerous applications (e.g., auto-completion).
Here, each infilling task masks one code token and asks the model to predict that missing token.
We aim to identify models performing accurately, prompt organizations leading to better results, and potential reasons for successful or failed infilling.
Thus, regarding TIB as positive cases, we evaluate three metrics:

\vspace{2pt}\noindent$\bullet$~\textbf{Specificity} (Consistency). On the TIB-free $D$, if the model predicts the same as the masked token, we consider it to have no TIB. High specificity indicates fewer false alarms. \\
\vspace{2pt}\noindent$\bullet$~\textbf{Recall} (Sensitivity). On the $D'$ of synthesized TIBs, if the predicted token differs from the masked, a potential TIB may exist. High recall indicates more detectable TIBs. \\
\vspace{2pt}\noindent$\bullet$~\textbf{Speed}. This measures the time required to complete the infilling task, reflecting scalability. \\

\begin{figure}
\center
\includegraphics[width=0.49\textwidth]{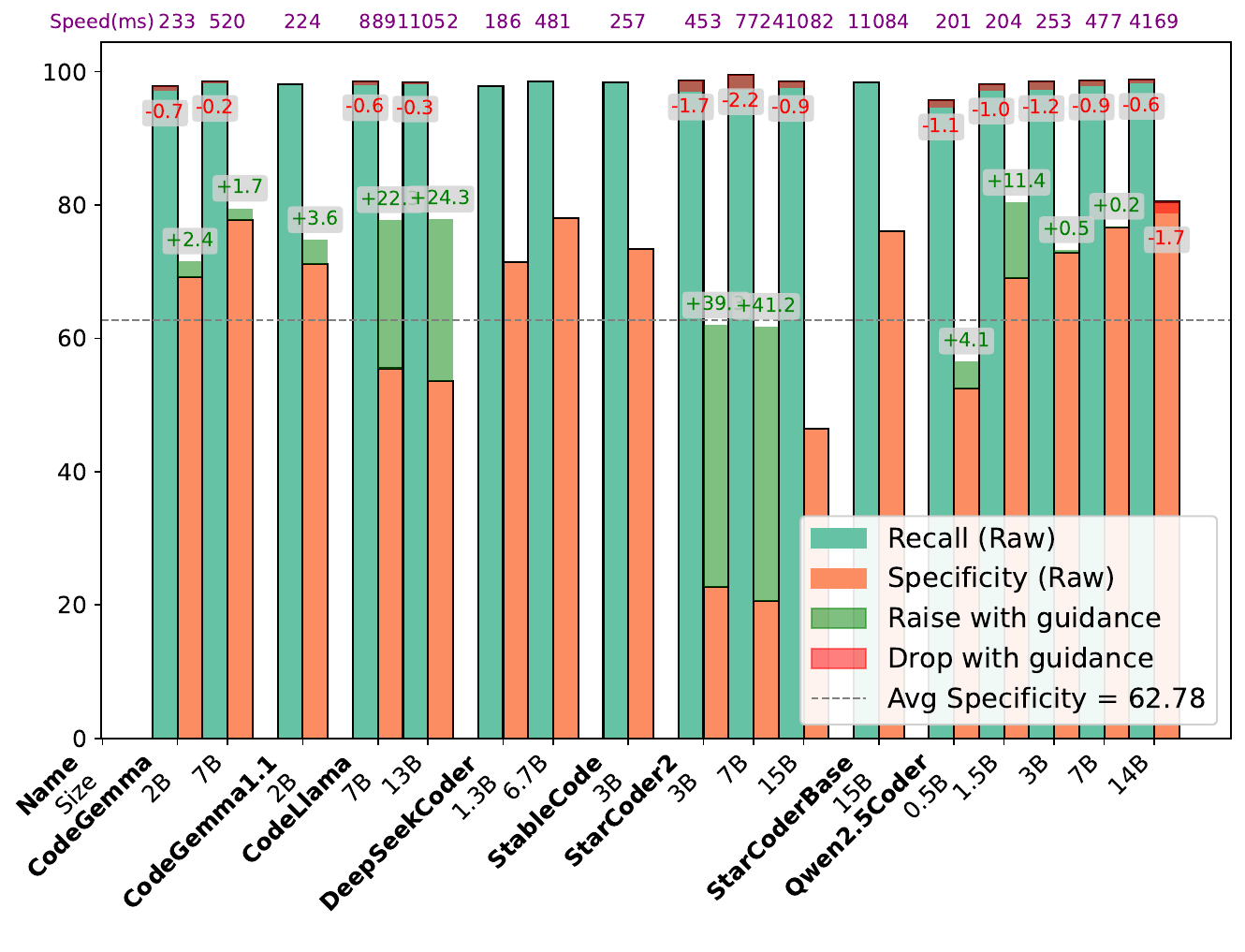}
\caption{Code-specific models' infilling performance.}
\label{fig:local-infilling}
\end{figure}

Similar to previous work~\cite{ahmadi2021finding}, we intentionally avoid using metrics like F1 score because the distribution of TIB varies (\eg on languages), and such distribution is unavailable.
\autoref{fig:local-infilling} summarizes the results.
Due to GPU memory constraints, models exceeding 15B are excluded from our evaluation.
While our H100 GPU with 80 GB memory can successfully load these models, the practical memory requirements extend beyond simple model loading.
The self-attention mechanism, fundamental to transformer inference, induces space-complexity quadratic in the input/output sequence length~\cite{keles2023computational}.
For example, the major memory requirements break down as follows for CodeLlama 13B model quantized at FP16: \mytilde24 GB for model weights ($13 \times 10^9 \times 2 ~\mathrm{bytes}$) and \mytilde48GB for attention layers of 4,000 tokens ($4000^2 \times 40~\mathrm{layers} \times 40~\mathrm{heads} \times 2~\mathrm{bytes}$).
Adding other components (\eg activation and KV-cache), inference with code length beyond 4,000 tokens could exhaust GPU memory on our system.
Therefore, cherry-picking the input context becomes crucial for efficiency.
We discuss potential countermeasures in \autoref{sec:discussion}.
That said, our solution design (\autoref{sec:design}) is orthogonal to specific models.


\parhead{Measurement Results}
Our findings reveal that models universally exhibit high recall (\mytilde 98\%).
However, specificity varies considerably.
StarCoder2 frequently fails to restore the original tokens, whereas DeepSeek, CodeGemma, and Qwen models achieve a specificity of 70–80\%.
Larger models in the same family generally exhibit better specificity and recall but experience nonlinear slowdowns due to GPU memory constraints when handling extended contexts.

A key factor affecting specificity is that models usually overlook the validity of infilling results.
Some models insert irrelevant text, such as comments or extraneous statements, because they are unaware of legitimate code token syntax.
Besides, function calls exhibit lower specificity than variables, operators, or literals.
We hypothesize that these results are from missing function definitions or external context, which models cannot infer from the snippet alone.

\parhead{Constrained Token Generation}
Some frameworks can impose constraints on generated tokens~\cite{willard2023efficientguidedgenerationlarge}.
We also integrated Guidance~\cite{guidance} into our experiments and denote the metric changes on the bars in \autoref{fig:local-infilling}.
By restricting generated tokens through regular expressions or lists of candidates, specificity significantly improves (especially for Code Llama and StarCoder2), albeit with a slight reduction in recall.
However, some models, like DeepSeek Coder, are incompatible with Guidance due to byte encoder errors.
Guidance also introduces additional runtime overhead.

\begin{table}[]
\centering
\fontsize{6.68pt}{8.68pt}\selectfont
\caption{Infilling performance under different contexts.}
\label{tab:infilling-context}
\fontsize{6.68pt}{8.68pt}\selectfont
\begin{tabular}{ll|cc|cc|cc}
\hline
\multicolumn{2}{c|}{Model}           & \multicolumn{2}{c|}{Function} & \multicolumn{2}{c|}{File} & \multicolumn{2}{c}{Sliced File} \\
Name and Size           & Length   & Spe.    & Time    & Spe.   & Time  & Spe.     & Time    \\ \hline
CodeGemma 2B   & 8K       & 71.2    & 441     & 19.8   & \textbf{5621}   & 56.9     & 1442     \\
StableCode    3B   & 16K      & 73.4    & \textbf{257}     & 68.5   & 10657 & 71.2     & \textbf{989}      \\
DeepSeek 6.7B & 16K      & \textbf{78.0}    & 481     & \textbf{85.8}*  & 9340   & \textbf{77.4}     & 1638     \\
Code Llama    13B  & 100K     & 53.6    & 11052    & 60.1* & 8355  & 57.3*    & 5075     \\
Qwen2.5  7B   & 32K      & 76.6    & 477      & 85.6  & 8773   & 75.1     & 11129    \\ \hline
\end{tabular}

* Some tasks are excluded due to runtime errors (e.g., insufficient GPU memory).
\end{table}

\parhead{Context Dependency}
We also explored different context organizations: 1) Function: only the function definition, which is used previously in \autoref{fig:local-infilling}; 2) File: the entire file; 3) Sliced File: function definition plus global variables and function declarations in the file.

\autoref{tab:infilling-context} shows that richer contexts can improve specificity by up to 9\% (e.g., DeepSeek Coder, Qwen2.5 Coder, Code Llama).
At the same time, the effect on recall is not significant and is omitted from the figure.
However, extensive context truncation degrades performance for smaller models like CodeGemma with stricter token length limits.
We also note that extended inputs incur slowdowns in inference, and may even raise runtime errors due to insufficient GPU memory (\eg Code Llama 13B).
Thus, context engineering should be paired with various models for better overall performance.

\parhead{Impact on Nearby Tokens}
TIB can negatively affect the infilling task in its nearby context.
When a TIB presents in a line, the specificity of the infilling tasks on other positions in the same line drops 6\%-7\%.
We further design a highlighting prompt in \autoref{sec:design:hl} to enhance the robustness of our pipeline.
A potential explanation is that TIBs increase the perplexity, leading to larger uncertainty in token prediction~\cite{tang2024science}.
This also implies that inconsistent infilling results conducted on single tokens may unveil bugs related to multiple code tokens.

\parhead{ Conclusion}
We highlight these conclusions from our study:

\noindent\textit{(1) High Recall but Mixed Specificity.}
Most models detect more than 97\% of synthesized TIBs, showing strong sensitivity to TIB.
This could further foster TIB detectors where only a few TIBs can evade.
However, many models produce 15–30\% inconsistencies on TIB-free samples, even with constrained token generation and rich contexts.
As no single model–prompt configuration consistently excels across both performance metrics, TIB detection should not rely on a single model.


\noindent\textit{(2) Scalability Limitations on TIB Detection.}
Despite their ability to filter out numerous true negatives, plenty of false alarms can be fired.
However, as TIBs in real-world code might account for less than 1\%, using a single model can result in poor precision.
Even though constrained token generation and richer context can enhance specificity and mitigate this concern, the enhancement is neither universal to arbitrary models nor enough for a scalable detector.
We need additional measures to overcome this concern.

\section{Design and Implementation of \name}
\label{sec:design}


Inspired by the measurement, we present \name, a system to detect TIBs effectively and efficiently in various programming languages.
Our objective is to build a TIB detector that is scalable, flexible, and language-agnostic.
Motivated by the examples and measurements, we believe that LLMs are well-suited for TIB detection, potentially outperforming other techniques for three reasons.
First, transformer-based models naturally handle context-dependent TIB.
Semantically incorrect code tokens, contextualized by surrounding code, typically have a lower probability, plus LLMs could ``understand'' code semantics.
Thus, TIBs are detectable via infilling- or prompt-based methods.
Second, LLM-based approaches are more flexible and general than specialized ML-based solutions, which often rely on strict assumptions.
For example, syntactically similar functions must exist in target codebases~\cite{ahmadi2021finding}.
Because LLMs are trained on massive multilingual corpora, they inherently support different languages.
Finally, LLMs can further assist developers in fixing bugs by offering interactive, natural-language explanations.

However, the measurement study (\autoref{sec:measurement}) also reveals shortcomings.
\sotalm and smaller code-specific models alone cannot reliably handle TIB detection at scale because of excessive cost and/or impractical accuracy.
Fortunately, Such weaknesses can be mitigated with the \textit{synergy of SOTA and lightweight models}, leveraging the strengths of each category.
The high expenses and FP rate of \sotalm can be contained if smaller and cheaper models eliminate most TIB-free tokens in advance at lower costs.
Meanwhile, suspicious locations identified by these smaller models can be verified further by \sotalm, reducing the distractions to \sotalm.
We also design new algorithms and adaptive prompts to facilitate TIB detection. 



\begin{figure}
\center
\includegraphics[width=0.49\textwidth]{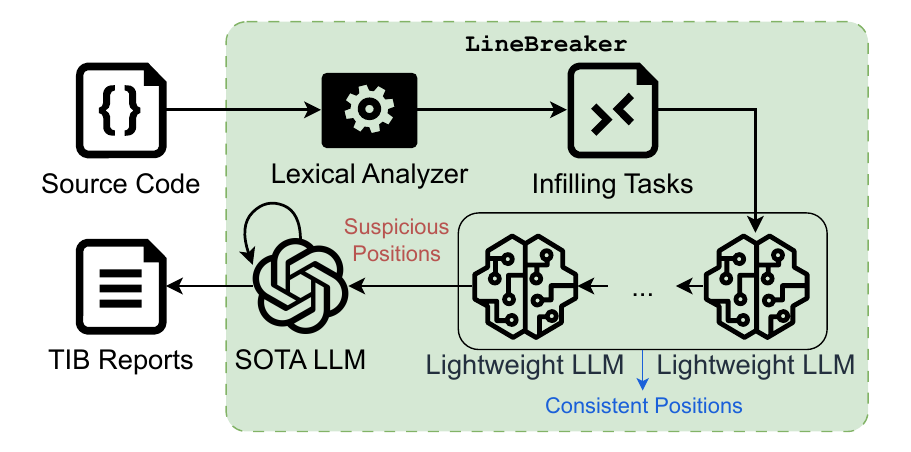}
\caption{Pipeline overview of \name.}
\label{fig:pipeline}
\end{figure}

\parhead{\name's Pipeline}
We depict \name's pipeline in \autoref{fig:pipeline}.
Initially, a lexical analyzer processes the code, compiling a comprehensive set of infilling tasks with code tokens being masked and token type information.
Then lightweight LLMs deployed locally conduct the infilling task, sequentially filtering out most TIB-free tasks and drastically reducing the search space.
It is worth mentioning that \name involves several different local models.
Next, following a carefully engineered prompt template, \name prompts \sotalm for further bug inspection.
\sotalm's responses undergo an automated filtration and are then aggregated into TIB reports for manual review.

By cascading models in this manner, \name significantly reduces false positives. 
Early stages remove most negatives before \sotalm is invoked, avoiding distractions and lowering costs (\autoref{sec:design:cascaded}).
Meanwhile, strategic token-generation control in the local models boosts specificity (\autoref{sec:design:controlled}), ensuring that most true negatives are discarded early.
Finally, \sotalm provides in-depth analysis of suspicious tokens, highlighting prompt engineering (\autoref{sec:design:hl}).
We evaluate the effectiveness of these new techniques later in \autoref{sec:eval:controlled}.

\subsection{Cascaded Detection}
\label{sec:design:cascaded}

The measurements in \autoref{sec:measurement} indicate that no single model can simultaneously deliver high recall, high specificity, and fast throughput for TIB detection.
\name employs a cascaded design that combines multiple models to balance these objectives.
We formalize our approach and discuss the rationale behind model selection.

\parhead{Problem Formalization}
Suppose our pipeline consists of $n$ stages, each equipped with a different language model.
Let $p_i$ denote the true negative rate (\ie TNR or specificity) of stage $i$,
$q_i$ the true positive rate  (\ie TPR or recall),
and $t_i$ the throughput (\ie execution speed).
Let $N_i$ represent the number of potential TIB cases to be processed at stage $i$,
and $\epsilon_i$ represents the ratio (\ie density) of actual TIB instances among $N_i$ after stage $i$, where $\epsilon_0$ being the initial TIB density before detection.
We can derive the number of true/false positives/negatives at each stage:
\begin{align*}
N_{TN, i} = N_ip_i(1-\epsilon_{i-1}) &\quad N_{TP, i} = N_iq_i\epsilon_{i-1} \\
N_{FP, i} = N_i(1-p_i)(1-\epsilon_{i-1}) &\quad N_{FN, i} = N_i(1-q_i)\epsilon_{i-1}
\end{align*}

After stage $i$, TN cases $N_{TN, i}$ are correctly excluded,
while FN cases $N_{FN, i}$ are TIBs missed at this stage.
TPs and FPs are escalated to the next stage, remaining $N_{i+1} = N_{TP, i} + N_{FP, i}$.
Thus, the total number of missed TIBs $M$ across the entire pipeline and the cumulative execution time $T$ are:
\begin{align*}
    M &= \sum_{i=1}^{n} N_i(1-q_i)\epsilon_{i-1} \quad \quad \quad
    T = \sum_{i=1}^{n} N_it_i
\end{align*}

Additionally, the TIB density after stage $i$ can also be derived:
\begin{align*}
    \epsilon_{i} &= \frac{N_{TP, i-1}}{N_{TP, i-1} + N_{FP, i}-1} = \frac{q_{i}\epsilon_{i-1}}{1-p_{i}-\epsilon_{i-1} + (p_{i}+q_{i})\epsilon_{i-1}}
\end{align*}

For an $n$-stage pipeline, the aggregated precision of all models before the final stage is then $\epsilon_{n-1}$.
The final stage adopts state-of-the-art \sotalm, which processes $N_{n-1}$ TIB candidates,
with the specificity $p_n$ and recall $q_n$.


We break down the cost into four factors for scalable TIB detection:
1) the cost of computing resources, $C_{comp}$ per unit time;
2) the cost of API invocations, $C_{api}$ per call for commercial LLMs;
3) the penalties associated with missed bugs, $C_{miss}$, due to potential financial loss;
and 4) manual inspection cost, $C_{check}$ per TIB.
We thus define the total cost
\begin{align*}
    C = C_{api}N_{n-1} + C_{comp}T + C_{miss}M + C_{check}N_{n}
\end{align*}

While the values of $C_{api}$ and $C_{comp}$ can be obtained from service providers,
$C_{miss}$ and $C_{check}$ are configurable.
This formula transforms the model selection problem into an optimization problem to minimize the total cost.
\ronghua{For an optimization problem, usually variables and constants should be specified explicitly. In another word, what factor do you change to minimize the optimization problem?}

\ronghua{Besides, I cannot see the role that such formalization and optimization play. How does this influence the final choice of the pipeline? There is no systematic approach to solve such an optimization problem so that we can reach the current design of the pipeline. Besides, is such a combination of models novel? I think machine learning works might have also done this.}

\begin{table}[]
\centering
\fontsize{6.68pt}{8.68pt}\selectfont
\caption{TIB detection cost under different parameters.}
\label{tab:formalization}
\fontsize{6.68pt}{6.68pt}\selectfont
\begin{tabular}{l|cccc|ccc}
\hline
\multirow{2}{*}{$\epsilon_0$}             & \multicolumn{4}{c|}{$pi=0.8$} & \multicolumn{3}{c}{$pi=0.6$} \\
            & $n=2$     & $n=3$     & $n=4$  & $n=5$    & $n=3$     & $n=4$ & $n=5$    \\ \hline
$10^{-2}$    & 365.4  & \textbf{218.6}  & 250.5 & 316.8 & 388.5  & \textbf{330.4}  & 351.7 \\
$10^{-3}$   & 295.7   & 76.4   & 38.6  & \textbf{37.1}  & 247.8   & 119.3  & \textbf{72.3}\\
$10^{-4}$  & 288.7   & 62.1   & 17.4   & \textbf{9.1}   & 233.7   & 98.2 & \textbf{44.4}\\ \hline
\end{tabular}
\end{table}

\parhead{\name's Pipeline Establishment}
First, we fix the constants in the cost formula.
We set $C_{miss}$ at \$500, which is preservative according to \cite{finifter2013empirical}.
Given the example in \autoref{fig:example1} that inspecting the report is eased by \sotalm, we set $C_{check}$ to \$2.
Based on the measurement (\autoref{sec:measurement:gpt}) and online data~\cite{lambda_gpu}, we set $C_{api}$ and $C_{comp}$ as \$0.025 per API call and \$2.49/hour.

Second, we discuss the number of optimal stages under an experimental configuration.
Fixing \sotalm as the final stage because its reasoning and natural-language feedback are instrumental for security experts.
We then simulate how many local lightweight LLMs should precede \sotalm.
TIB density $\epsilon_0$ varies concerning the dataset (\eg of different languages), so we pick representative values according to previous findings, where the number of bugs per 1,000 lines of code ranges from 0.1 to 25~\cite{mcconnell2004code}.
Borrowing average $q_i = 0.98$, selecting average and best $p_i$ as $0.6$ and $0.8$ from \autoref{fig:local-infilling}, we calculate the cost of our pipeline in \autoref{tab:formalization}.
We observe that
1) cascading more does not always derive cost-efficiency because missing bug penalties accumulate, and the marginal benefit shrinks;
2) enhancing specificity reduces the cost significantly;
3) positioning the faster model further to the SOTA reduces computation cost, although local model permutation is irrelevant to the overall accuracy.
These findings motivate us to construct better algorithms on token generation with enhanced specificity, and guide our model selection(see \autoref{sec:eval:realworld}).

\subsection{Controlled Token Generation}
\label{sec:design:controlled}
As measured in \autoref{sec:measurement:infilling}, local LLMs generate tokens in an unregulated manner, leading to many incorrect infilling results (\eg generating a docstring where a variable is expected).

Decoder-only models like Code Llama do not constrain which \textit{type} of code token to generate.
They blindly select most probable tokens sequentially until a special end-of-sequence symbol or length limit is reached.
A naive approach might check if the original token ranks high in probability.
However, once the model deviates from the correct token at any intermediate step, subsequent tokens are conditioned on the wrong prefix, rendering the original token improbable.
For example, when a variable name is masked but the model begins to generate documentation (\eg starting with \texttt{"""}), subsequent generation follows up this pattern rather than the expected variable name.
Existing frameworks like guidance can help but fail to constrain the generation of specific models.
Our solution is to inspect the probabilities of generated tokens and apply syntactic constraints on the output.



The high-level idea is to select tokens that adhere to the original code tokens in each iteration and inspect the probabilities to judge consistency.
As shown in \autoref{fig:pipeline}, the lexical analyzer aids infilling tasks construction with syntactical knowledge.
It inserts special tokens indicating the infilling position in each function, as shown in \autoref{fig:tokengen}.
Auxiliary information including the original tokens, token type (\eg variable and literal) is also recorded to constrain the generation in the subsequent steps.
Thus, \name selects each generated token by verifying whether it remains syntactically valid and aligns with the original code token.
For example, valid variable names in Python can only contain letters, underscores, and non-beginning digits.

\SetKwInput{KwInput}{Input}
\SetKwInput{KwOutput}{Output}
\SetKwData{KwIn}{in}
\SetKwComment{Comment}{/* }{ */}

\begin{algorithm}[t]
\fontsize{6.68pt}{8.68pt}\selectfont
\caption{Decoder-only Model Consistency Check}\label{alg:codellama-consistency}
\KwInput{
    maskedCode,
    originalTokens,
    probThresh,
    rankThresh,
}
\KwOutput{consistencyFlag}
\BlankLine
rankSum $\gets$  0; \\
generatedTokens, leftTokens $\gets$ [], originalTokens; \\
\While{leftTokens.length \textgreater 0}{
    nextTokenProbList  $\gets$ top\_k(CodeLlamaPredict(maskedCode, generatedTokens)); \\
    \ForEach{token, prob \textbf{in} nextTokenProbList}{
        \eIf{validateToken(token, typeof(token)}{
            \eIf{leftTokens.startswith(token)}{
                leftTokens.remove(token) \\
                generatedTokens.append(Token) \\
                \textbf{break to generate the next token};
            }{
                \If{prob \textgreater probThresh}{
                    \Return False;\\
                }
                rankSum $\gets$ rankSum + 1; \\
            }
        }{
            \textbf{continue to the next iteration};
        }
    }
    \If{rankSum \textgreater rankThresh}{
    \Return False;\\
    }
}
\Return True;\\
\end{algorithm}


\autoref{alg:codellama-consistency} inspects the probabilities of the generated candidates in each step.
For each generated \textit{token}, \name first checks if the token is syntactically valid.
Then, if \textit{leftTokens} starts with the generated \textit{token} (\ie the generation follows the original code token), it updates \textit{leftTokens} and continues the next generation.
Otherwise, this generation is penalized for not being consistent with the code token.
In cases where the probability \textit{prob} is greater than a threshold \textit{probThresh}, meaning there is another, more probable, and valid token that can deviate from the generation, the check directly returns \textit{False}.
Nevertheless, when deviation happens but \textit{prob} is less than \textit{probThresh}, it adds one to \textit{rankSum}, penalizing the original code token for not being the most probable generated \textit{token}.
Finally, consistency is determined by comparing \textit{rankSum} with a threshold.



\subsection{Highlighting Inconsistent Positions}
\label{sec:design:hl}

As observed in \autoref{sec:measurement:gpt}, \sotalm can produce many FN cases, especially with lengthy code snippets that distract the model with irrelevant context~\cite{sun2024llm4vuln}.
Thus, we design a prompt engineering technique emphasizing the suspicious tokens identified by lightweight LLMs.
Based on the most effective prompt template we measured, \name instructs \sotalm: ``Also, pay additional attention to these lines: \texttt{\{suspicious\_lines\}}'' \hongbo{update}.
Rather than the suspicious token, the whole line is highlighted because of our findings in \autoref{sec:measurement:infilling}, which indicate that a TIB at one position can increase perplexity for nearby tokens on the same line. 
Thus, inconsistencies may also imply TIBs at another position in the same line.




\subsection{Implementation}
\label{sec:design:implementation}

We implement \name in Python with roughly 7K lines of code.
It relies on tree-sitter~\cite{TreeSitter} to preprocess the source code, as it can parse multiple programming languages.
\name sources models from Hugging Face, and utilizes the transformer library to generate tokens.
\autoref{alg:codellama-consistency} is based on PyTorch framework.
\name also relies on OpenAI's official library to invoke SOTA LLMs deployed on the cloud.
Besides Python integration, we support the C language with 500 additional lines to demonstrate generality.

\section{Evaluation}
\label{sec:evaluation}

In this section, we first conduct controlled experiments on our prepared datasets (\autoref{sec:measurement:dataset}).
This is to (1) evaluate the effectiveness of individual components in our design
and (2) compare \name with existing bug detectors.
After that, we evaluate \name on real-world code repositories and demonstrate its ability to discover new bugs.
The experiment uses the same platform as in \autoref{sec:measurement:infilling}.

\subsection{Controlled Experiments}
\label{sec:eval:controlled}

We continue using dataset $D$ (TIB-free) and $D'$ (synthesized TIBs), measuring recall and specificity as in \autoref{sec:measurement}.
The evaluation also helps us concretize some parameters,
which will be used in a scalable real-world TIB detection in \autoref{sec:eval:realworld}.

\begin{table}[]
\centering
\caption{Performance of Selected Models with \autoref{alg:codellama-consistency}.\hongbo{include baseline?}}
\label{tab:codellama-params}
\fontsize{6.68pt}{8.68pt}\selectfont
\begin{tabular}{l|cc|
>{\columncolor[HTML]{C0C0C0}}c 
>{\columncolor[HTML]{C0C0C0}}c |cc|c}
\hline
\multirow{2}{*}{Model}                    & \multicolumn{2}{c|}{\textit{rankT}=0} & \multicolumn{2}{c|}{\cellcolor[HTML]{C0C0C0}\textit{rankT}=1} & \multicolumn{2}{c|}{\textit{rankT}=2} & \multirow{2}{*}{Time}      \\
               & Rec.          & Spe.         & Rec.                      & Spe.                     & Rec.          & Spe.         &  \\ \hline
StarCoder2 3B        & 96.7          & 74.6         & 91.0{\textcolor{red}{$\downarrow$}}                       & 82.4{\textcolor{green}{$\uparrow$}}                       & 87.1          & 85.2         & 727   \\
Code Llama 7B        & 97.5          & 65.8         & 92.2{\textcolor{red}{$\downarrow$}}                      & 75.3{\textcolor{red}{$\downarrow$}}                      & 89.6          & 78.5         & 1601  \\
DeepSeek Coder 6.7B* & 97.4          & 79.7         & 92.3{\textcolor{red}{$\downarrow$}}                      & 86.6{\textcolor{green}{$\uparrow$}}                     & 89.3          & 89.2         & 1606  \\ \hline
\end{tabular}

* Guidance does not support DeepSeek models, so the metrics are compared with Raw setup. Others are compared with guidance in \autoref{fig:local-infilling}. \textit{rankT} referes to \textit{rankThreshold}.
\end{table}

\parhead{Controlled Token Generation Algorithm}
Next, we evaluate \autoref{alg:codellama-consistency} on decoder-only models.
We evaluate \autoref{sec:measurement:infilling} again with different \textit{rankThresh} (\textit{rankT}) on decoder-only models, and present the results of selected models in \autoref{tab:codellama-params}.
We set \textit{probThresh} = 0.9 after observing minor variations in performance at different probability thresholds.
Our algorithm enables constrained token generation for models not supported by guidance, such as DeepSeek Coder.
While our algorithm incurs a modest speed penalty, it notably boosts specificity that offsets minor recall loss, especially when \textit{rankthresh} $\leq 1$.
The recall and specificity of our algorithm are at least comparable to guidance with selected parameters.
Since TIBs are rare in real-world code, we choose \textit{rankThresh} $=1$, to reduce false positives and improve overall precision.


\parhead{Highlighting}
To measure the effectiveness of the highlighting technique (\autoref{sec:design:hl}), we integrate it into the best prompt template.
The results are listed in \autoref{tab:gpt4-templates} along with the results in measurement \autoref{sec:measurement:gpt}.
The highlighted lines also include up to four randomly selected lines in each sample to simulate false positives propagated from the previous models.
Considering both specificity and recall, highlighting drastically enhances the performance.
This is because GPT-4 refocuses on the suspicious line, mitigating the distraction problems.
Importantly, highlighting is only feasible when local models have already identified specific suspicious tokens.




    
    

\subsection{Real-world TIB Detection}
\label{sec:eval:realworld}

To demonstrate its real-world impact, we evaluate \name on Python and C repositories on Github, which are (1) popular (with 1k+ stars), and (2) actively maintained.
The results are summarized in \autoref{tab:breakdown}.

\noindent\textbf{New Bugs.}
\name successfully found 123 new TIBs, we have submitted 69 bug fixes as pull requests, of which 41 have been merged or confirmed by developers.

\noindent\textbf{Precision and Practicality.}
\name achieves a bug detection precision of 23.5\%-36.3\%, significantly outperforming existing semantic bug detection works~\cite{ahmadi2021finding} on real-world projects (12.0\%).
We believe this precision is also practical, especially for developers who prioritize security, as suggested by the recent study~\cite{ami2023false}.
Moreover, LLM's detailed and intuitive explanations make our bug report easy to inspect.
Actually, each report takes only about \emph{one minute} for our team members to review. 
In other words, by inspecting \name's output, one can \emph{find one true TIB roughly every four minutes}.


\parhead{Model Selection}
As we conduct the evaluation on popular repositories, we assume $\epsilon_0$ ranges around $10^{-3}$~\cite{mcconnell2004code}.
Thus, our pipeline consists of three local models plus GPT-4o according to \autoref{sec:design:cascaded}.
Such a configuration misses fewer TIBs with additional but acceptable human inspection cost and slight accuracy reduction.
The optimal selection of three local models is Qwen2.5 Coder 1.5B, CodeGemma 7B, and Qwen2.5 14B.
However, as fewer models were available when we conducted this evaluation, we utilized StableCode and CodeLlama models.
GPT-4o was used as the state-of-the-art LLM to inspect the filtered results.
We use the prompt template shown in \autoref{fig:prompt-12FMCa} plus suspicious lines highlighting.
New models such as Llama 3 have been recently released.
However, their APIs previously offered limited support for JSON output, which makes it difficult to integrate into an automated detection pipeline.
We leave the measurement and integration of these models as future work.

We thus estimate \name's end-to-end performance using the formulas in \autoref{sec:design:cascaded}, deriving a theoretical accuracy of 10.5\%.
Such value is comparable to our evaluations in \autoref{sec:eval:realworld}.
Notably, relying solely on GPT-4 as a TIB detector yielded a precision of only 0.7\%.
In contrast, \name effectively amplifies the density of TIB by 105 times, making scalable TIB detection and manual inspection practical.


\begin{table}[]
\centering
\fontsize{6.68pt}{8.68pt}\selectfont
\caption{Summary of real-world TIB detection results.}
\label{tab:breakdown}
\begin{tabular}{l|l|l}
\hline
                             & \textbf{Python Repos} & \textbf{C Repos} \\ \hline
 Repositories            &       80                &          74            \\ 
\# Submitted PR    &     55                  &            14          \\ 
\textbf{\# Merged / Confirmed PR}   &      32                 &            9          \\ \hline \hline
\# Functionality Bugs &       37                &          7            \\ 
\# Security Bugs      &       8                &           3           \\ 
\# Bad Smell        &        48               &           20           \\ \hline \hline
\# Inconsis. Token after Local models    &     2,679                  &      733          \\ 
\# Reports from GPT-4    &     1,082                  &      473                \\ 
\# Reports after Filtering    &     314                  &   77                   \\ 
\# Correct Reports (may contain $\ge 1$ bugs)    &     74                  &    28                  \\ \hline
\textbf{Precision (\%)}       &  23.57\%                     &     36.36\%                 \\ \hline 
\end{tabular}
\end{table}

\parhead{Detecting TIBs in Real-world Repositories}
\name processed 80 popular Python repositories from GitHub.
Initially, 80K infilling tasks are derived from static analysis.
After dropping the consistent ones and tasks associated with very long code snippets (\eg configuration and data files), 2,679 inconsistencies are left after local model filtering.
Escalating these functions containing inconsistencies to GPT-4 for TIB detection, 1,082 reports remained, of which 314 persisted after filtering based on the properties mentioned in \autoref{sec:measurement:gpt}.
Three security experts spent 12 man-hours totally inspecting these reports, confirming that 74 reports are true positives.
This resulted in a precision of 23.57\% on Python repositories.
The detection requires less than 20 machine hours in total and less than \$40 to query GPT-4o APIs.
With a server cost of \$52, the total cost to find these TIBs is less than \$100.

We also experimentally test \name on C code to demonstrate its generality.
It generates infilling tasks according to abstract syntax tree AST similarity strategies~\cite{baxter1998clone,jiang2007deckard} following a previous research~\cite{ahmadi2021finding}.
We observe an enhanced accuracy of \name on C repositories, 36.36\%.
A potential reason is that Code Llama performs significantly better on C, filtering out 97.1\% consistent tokens.
A potential explanation is that C language induces fewer defects than Python~\cite{ray2014large}, making masked tokens more predictable.
Besides, we prototype support for the Rust programming language and experimentally discover one TIB in a repository.
These evaluations showcase the generality of \name across various languages, backing its scalable deployment.

\parhead{Repository Exclusion}
We intentionally exclude 10 repositories from our dataset due to their negative impact on LLM's performance:
(a) Projects written in outdated languages (\eg Python 2), which could confuse LLMs as they are trained on a more recent code corpus;
(b) AI-related projects (\eg chat-langchain~\cite{github:chat-langchain}).
We observed downgraded LLM performance for them, likely because their fast-evolving AI techniques surpass LLMs' training corpus;
(c) Security-related or low-level system projects (\eg hosts~\cite{github:hosts}), whose special or unusual functionalities often lead to reduced specificity.
For example, exploit generators intentionally include insecure code, rendering respective infilling tasks regarded as inconsistent by local models, or reported as vulnerable by GPT-4.
We believe such repositories could be better handled with fine-grained prompt engineering, which we leave as future work.

\subsection{Study of the Discovered TIBs}
\label{sec:eval-new-TIBs}

\name successfully identifies multiple real-world TIBs in impactful software projects,
some of which lead to severe security consequences like memory corruption and denial-of-service attacks.
Notably, a director of wolfSSL~\cite{wolfSSL} - a popular and security-critical library - scheduled a meeting with us to discuss our found vulnerabilities and fixes.
We showcase some TIBs we found in this section.

\begin{lstlisting}[
    language=C,
    basicstyle=\fontsize{6.68pt}{8.68pt}\selectfont\ttfamily,
    escapeinside={|}{|}
]
switch (optionId) { ...
case InterpreterThreadFrameStackSize:
  return s_threadFrameStackSize;
case InterpreterThreadExceptionFlowSize:
  return |\colorbox{red!30}{\sout{s\_threadFrameStackSize}}||\colorbox{green!50}{s\_threadExceptionFlowSize}|;}
\end{lstlisting}
\noindent\textit{Case 1.}
The second switch-case in the above code snippet wrongly returns the same variable as the first one,
where the correct identifier should be \texttt{s\_threadExceptionFlowSize}.
Since the returned size here will be used for buffer initialization,
buffer overflow can happen later, causing severe security consequences such as privilege escalation.
We suspect the developer forgot to change the variable name when copying and pasting the return statement.


\noindent\textit{Case 2.}
\autoref{code:method-misuse} shows a TIB in the security-critical WolfSSL library.
The \texttt{sizeof()} function is redundantly applied to \texttt{xSize}, which is already a size-representing variable.
This leads to an unexpectedly smaller buffer size, likely causing overflows with severe security consequences.
We suspect this is caused by variable naming confusion.


\begin{lstlisting}[
    language=Python,
    basicstyle=\fontsize{6.68pt}{8.68pt}\selectfont\ttfamily,
    escapeinside={|}{|}
]
if has_aux:
    example, *aux = example
else:
    aux = tuple()
rand_state = random.Random(|\colorbox{red!30}{\sout{aux[-1]}}||\colorbox{green!50}{aux[-1] if aux else 0}|)
\end{lstlisting}
\noindent\textit{Case 3.}
\texttt{aux} could be an empty tuple in the above Python code snippet,
where accessing its last element with \texttt{aux[-1]} will cause an index out of range exception, leading to DoS.


\subsection{Comparison with Previous Work}
\label{sec:eval:comparison}
TIB covers semantic bugs like variable misuse, which are also targeted by some existing works.
However, direct comparisons with them are often difficult due to reasons like artifact unavailability~\cite{vasic2018neural}, high training costs~\cite{allamanis2021self}, and different targeted languages~\cite{allamanis2018learning, dinella2020hoppity, pradel2018deepbugs}.
Therefore, we compare with some representative works with thorough design-level analysis, with auxiliary experiments when applicable.

\noindent\textbf{Pre-LLM Approaches.}
Before LLMs' flourish, machine learning based approaches relied on limited and dedicated datasets for model training, affecting flexibility and generality.
DeepBugs~\cite{pradel2018deepbugs} targets only three specific types of name-based bugs with known patterns, limiting its generality.
FICS~\cite{ahmadi2021finding} recognizes outlier functions as potential bugs, with AST-based embeddings.
However, semantic information (\eg variable names) is not included in such embeddings, rendering it ineffective in TIB detection.
Furthermore, FICs assumes the existence of a ``correct'' function in the codebase to spot the similar but buggy one, which may not hold especially in smaller codebases.
Unlike these systems, \name requires no training, fine-tuning, or strong assumptions on bug patterns, codebase, or programming languages.

\noindent\textbf{LLM-Based Approaches.}
FLAG~\cite{flag} utilizes LLM for semantic bug detection at the line level and is the closest related work. 
It compares the original code against the code generated by LLM line by line, given other lines as the prompt in each line's iteration. 
To ensure a fair comparison, considering FLAG's high economic costs and line-by-line detection, we evaluated FLAG on 200 evenly sampled functions from $D$ and $D'$ and present the line-level metrics in \autoref{tab:flag_comparison}.
Among these, 100 samples/lines contain TIB (\ie 100 positive line-level cases), while the remaining lines constitute negative instances.
The results show \name significantly outperforms FLAG in all essential metrics: precision (0.9\% $\rightarrow$ 14.3\%), recall (32\% $\rightarrow$ 69\%), and monetary cost on LLM APIs.
%

\begin{table}[t]
\centering
\fontsize{6.68pt}{8.68pt}\selectfont
\caption{Line-level metrics comparison with FLAG on the testset.}
\label{tab:flag_comparison}
\begin{tabular}{l|cc|cccc|c}
\hline
\textbf{} & Precision & Recall & $\text{TP}_L$ & $\text{FP}_L$ & $\text{TN}_L$ & $\text{FN}_L$ & Cost \\ \hline
FLAG & 0.9\% & 32.0\% & 32 & 3732 & 5396 & 68 & 37.4\$ \\ \hline
\name & \textbf{14.3\%} & \textbf{69.0\%} & 69 & 413 & 8715 & 31 & 5.1\$ \\ \hline
\end{tabular}
\end{table}

%
%
%
%
As seen, FLAG has many false alarms and misses the majority of TIBs, yet with high API cost (\mytilde \$0.19 per function).
We avoid timing the executions because both designs permit extensive concurrency optimization as computation power and API rate limit allow.
%
\name's superior performance roots in its effective pipeline design. 
It combines efficient yet capable local LLMs with the SOTA LLM, boosting the performance while maintaining a lower cost. 
%
%
%
Additionally, FLAG's cost increases quadratically with the number of lines.
Although \name induces expense on GPU hours because of running models locally, its API cost is then linear in the number of lines when using the SOTA LLM.

\section{Related Work}
\label{sec:related}




\parhead{Semantic Bug Detection}
Semantic bug detection is an active research area.
Besides comparative work introduced in \autoref{sec:eval:comparison}, various methodologies address bugs exhibiting well-defined and ambiguous patterns. 
Li et al. pioneered the application of LLMs to target Use Before Initialization bugs, demonstrating LLMs' potential to enhance conventional static analysis techniques \cite{li2024enhancing}. 
Further advancements by researchers like Sun et al. and Wei et al. have expanded LLM applications to address a wider range of logic bugs, including API misuse \cite{sun2024llm4vuln, sun2024gptscan, wei2024demystifying, xia2024exploring, liu2025generating, yang2025midas}. 
Traditional methods such as static analysis and formal verification also play significant roles in identifying and resolving semantic bugs with fixed patterns \cite{liu2019avatar, huang2017pallas, fehnker2013model, li2021mirchecker, nguyen2013semfix}. 
Recently, Natural Language Processing (NLP) has been leveraged to analyze documentation and pinpoint potential semantic inconsistencies or discrepancies between implementation and documentation \cite{chen2019devils, panthaplackel2021deep, chen2021bookworm, chen2022seeing, chen2023sherlock}. 
However, these NLP-based approaches heavily rely on the quality of the documentation, facing substantial challenges with ambiguous or outdated materials prevalent in open-source environments.

\ignore{
FLAG~\cite{flag} is a recent initiative that utilizes LLMs for bug detection by masking every line of code and comparing the code line generated by ChatGPT with the original line.
As we demonstrated in \autoref{sec:eval:controlled}, such methods exhibit limitations when applied to large-scale, real-world analysis.
Beyond LLM-based methods, other approaches such as~\cite{ahmadi2021finding, pradel2018deepbugs}  utilize learning-based methods to pinpoint potential bugs.
However, these methods require specific code characteristics or depend on ideal, comprehensive training datasets.
}

\parhead{LLM for Software Engineering}
Recent advancements have made LLMs for coding tasks widely accessible through public API or open-source sharing of code and models.
LLMs have demonstrated significant contributions to areas like fuzzing~\cite{li2024exploring, Meng2024LargeLM, xia2024fuzz4all}, code repair~\cite{xia2023automated, Joshi2022RepairIN, Fan2022AutomatedRO, Pearce2021ExaminingZV, Jin2023InferFixEP}, test generation~\cite{yang2024evaluation, chen2024chatunitest}, code-comment consistency checking~\cite{zhang2024leveraging}, etc.
Despite these developments, applying LLMs in scalable bug detection, specifically TIB, remains underexplored and largely experimental.


        


\section{Discussion and Conclusion}
\label{sec:discussion}

\parhead{Limitations}
Decoder-only models like Code Llama are currently limited to infilling at a single position each time. 
Although \name independently verifies the consistency of each code token, it may not detect sophisticated bugs associated with multiple discrete positions.
Additionally, LLMs sometimes fall short when dealing with corner cases, especially when the correct code appears counterintuitive.
This snippet demonstrates an FP case involving the transmission of a firmware header.
\begin{lstlisting}[
    language=C++,
    basicstyle=\fontsize{6.68pt}{8.68pt}\selectfont\ttfamily,
    escapeinside={|}{|},
    breaklines=true,
    breakatwhitespace=false % Equivalent to "breakanywhere"
]
if (!securedSend(0x00, 128, (const uint8_t *)fwData->getBytesNoCopy())) ...
if (!securedSend(0x03, 256, (const uint8_t *)fwData->getBytesNoCopy() + 128)) ...
if (!securedSend(0x02, 256, (const uint8_t *)fwData->getBytesNoCopy() + (|\fcolorbox{red}{white}{388}|))) ...
\end{lstlisting}
In the third \texttt{if} statement, GPT-4 believes the correct offset is 384 (128+256) instead of 388.
It assumes that since 384 bytes of data have already been sent, 384 should be the offset for the next chunk.
Relevant documentation is missing from the code repository.
Since software development heavily relies on external libraries and standard specifications, LLMs can make mistakes when related information is missing from the context (\eg proprietary code).
This drawback might be mitigated by compensating for richer context~\cite{ahmed2024studying}.
For example, RAG~\cite{lewis2020retrieval, lu-etal-2022-reacc} can intelligently fetch external knowledge.

\parhead{GPU Memory Constraint}
Our measurement experiments on local large code models (\autoref{sec:measurement:infilling}) revealed that inference tasks may fail with extended context.
The direct solution is running \name on a system with connected GPUs that offer a larger GPU memory pool.
However, some optimizations can reduce the memory consumption.
For example, FlashAttention~\cite{dao2022flashattention} reduces the space complexity linear in the context length, but requires hardware support.
Quantization~\cite{lin2024awq} compresses the model weights by reducing parameters' numerical precision, also leading to less memory usage.

\parhead{Future Work}
\name heavily relies on existing infilling models to mark suspicious TIBs. 
It may be beneficial to fine-tune or train a large model with improved specificity and recall. 
The fine-tuned model demonstrates better capability for auditing smart contracts~\cite {ma2024combining}.
Although simple, replaced tokens detection~\cite{clark2020electra} fits our scenario with a new model trained. 
Researchers currently consider this task as a form of pretraining, which enhances the model’s capability to comprehend semantics more effectively.

Another potential engineering direction is in the context, which serves as a foundation for inference.
Our measurement study explored contexts at coarse-grained levels, such as files and functions (\autoref{sec:measurement:infilling}).
Although out of \name's scope, its design excludes bugs spanning multiple positions in the program.
Fine-grained context engineering may leverage code slicing~\cite{slicing1,slicing2}, where relevant functions from different files can be curated, and irrelevant secure code lines can be removed.
Thus, more complex bug patterns involving tokens scattered at different places might be captured.

\ignore{
As mentioned in \autoref{sec:measurement:infilling}, TIB can negatively affect the consistency of infilling tasks on surrounding positions, which might be qualitatively explainable with a confidence measure~\cite{spiess2024calibration}.
Enhancing this observation and providing a security guarantee for TIB detection is also worth future investigation.
}
    


\section{Conclusion}
\label{sec:conclusion}
We systematically measured LLMs' capabilities to detect TIBs and identified their strengths and weaknesses.
Based on the measurement, we design \name, a TIB detector using LLMs.
It discovered 123 unknown bugs, demonstrating its effectiveness, scalability, and generality.

\section{Acknowledgement}
\label{sec:ack}
We sincerely thank Yuhui Hong, the anonymous reviewers, and the shepherd for their valuable insights.
This work was supported in part by the National Science Foundation under Grant No. 2154199, and Lilly Endowment, Inc., through its support for the Indiana University Pervasive Technology Institute.
Any opinions, findings, conclusions, or recommendations expressed in this material are those of the
author(s) and do not necessarily reflect the views of the funding agencies.

\bibliographystyle{plain}
\bibliography{ref}

\appendix

\section{Detailed Measurement Results}

\section{Prompt Templates}
\label{sec:templates}

\begin{figure}[H]
\begin{tcolorbox}[colback=gray!10!white, 
                  colframe=black, 
                  fonttitle=\small,
                  fontupper=\small,
                  fontlower=\small,
                  arc=2mm, 
                  boxsep=5pt, 
                  left=0pt, 
                  right=0pt, 
                  top=-2pt,
                  bottom=-2pt,
                  boxrule=0.5pt 
                  ]
\textbf{Round 1 Properties}: \texttt{code\_line} + \texttt{explanation} \\
\textbf{Round 1 Prompt}: \texttt{\{code\}}
Output exact lines of semantic bugs and concise explanations of the bugs.
\end{tcolorbox}
\caption{Prompt template 1.}
\label{fig:template_1}
\end{figure}

\begin{figure}[H]
\begin{tcolorbox}[colback=gray!10!white, 
                  colframe=black, 
                  fonttitle=\small,
                  fontupper=\small,
                  fontlower=\small,
                  arc=2mm, 
                  boxsep=5pt, 
                  left=0pt, 
                  right=0pt, 
                  top=-2pt,
                  bottom=-2pt,
                  boxrule=0.5pt 
                  ]
\textbf{Round 1 Properties}: \texttt{code\_line} + \texttt{explanation} + \texttt{fixed\_line} + \texttt{token\_level} \\
\textbf{Round 1 Prompt}: \texttt{\{code\}}
Output exact lines of semantic bugs, concise explanations of the bugs, the fixed line of the bugs, and if the bugs are related to single or a few tokens.
\end{tcolorbox}
\caption{Prompt template 1FT.}
\label{fig:template_1FM}
\end{figure}

\begin{figure}[H]
\begin{tcolorbox}[colback=gray!10!white, 
                  colframe=black, 
                  fonttitle=\small,
                  fontupper=\small,
                  fontlower=\small,
                  arc=2mm, 
                  boxsep=5pt, 
                  left=0pt, 
                  right=0pt, 
                  top=-2pt,
                  bottom=-2pt,
                  boxrule=0.5pt 
                  ]
\textbf{Round 1 Properties}: \texttt{code\_line} + \texttt{explanation} \\
\textbf{Round 1 Prompt}: \texttt{\{code\}}
Output exact lines of semantic bugs, concise explanations of the bugs, the fixed line of the bugs, and if the bugs can be fixed by merely changing a variable/method name. \\
\textbf{Round 2 Properties}: \texttt{code\_line*} + \texttt{explanation*} + \texttt{fixed\_line} + \texttt{token\_level} \\
\textbf{Round 2 Prompt}:
Inspect these bugs, excluding 1. incorrect or unlikely bugs; 2. non-semantic bugs. Check the left bugs only break the intended functionality or lead to vulnerabilities. Answer if the bugs are related to single or a few tokens.
\end{tcolorbox}
\caption{Prompt template 1/2FT.}
\label{fig:template_12FM}
\end{figure}

\begin{figure}[H]
\begin{tcolorbox}[colback=gray!10!white, 
                  colframe=black, 
                  fonttitle=\small,
                  fontupper=\small,
                  fontlower=\small,
                  arc=2mm, 
                  boxsep=5pt, 
                  left=0pt, 
                  right=0pt, 
                  top=-2pt,
                  bottom=-2pt,
                  boxrule=0.5pt 
                  ]
\textbf{Round 1 Properties}: \texttt{code\_line} + \texttt{explanation} \\
\textbf{Round 1 Prompt}: \texttt{\{code\}}
Output exact lines of semantic bugs, concise explanations of the bugs, the fixed line of the bugs, and if the bugs can be fixed by merely changing a variable/method name. \\
\textbf{Round 2 Properties}: \texttt{code\_line*} + \texttt{explanation*} + \texttt{fixed\_line} + \texttt{token\_level} \\
\textbf{Round 2 Prompt}:
Inspect these bugs, excluding 1. incorrect or unlikely bugs; 2. non-semantic bugs. Check the left bugs only break the intended functionality or lead to vulnerabilities. Answer if the bugs are related to single or a few tokens. \\
\textbf{Round 3 Properties}: \texttt{code\_line*} + \texttt{explanation*} + \texttt{priority} \\
\textbf{Round 3 Prompt}:
Now, Evaluate the priority of detected bugs in previous round. Assign higher priority for logic bugs, security vulnerabilities, or unexpected functionality, and lower priority for bad coding styles (bad smells).
\end{tcolorbox}
\caption{Prompt template 1/2FT/3P.}
\label{fig:template_12FM3P}
\end{figure}

\begin{figure}[H]
\begin{tcolorbox}[colback=gray!10!white, 
                  colframe=black, 
                  fonttitle=\small,
                  fontupper=\small,
                  fontlower=\small,
                  arc=2mm, 
                  boxsep=5pt, 
                  left=0pt, 
                  right=0pt, 
                  top=-2pt,
                  bottom=-2pt,
                  boxrule=0.5pt 
                  ]
\textbf{Round 1 Properties}: \texttt{code\_line} + \texttt{explanation} \\
\textbf{Round 1 Prompt}: \texttt{\{code\}}
Output exact lines of semantic bugs, concise explanations of the bugs, the fixed line of the bugs, and if the bugs can be fixed by merely changing a variable/method name. \\
\textbf{Round 2 Properties}: \texttt{code\_line*} + \texttt{explanation*} + \texttt{fixed\_line} + \texttt{token\_level} \\
\textbf{Round 2 Prompt}:
Inspect these bugs, excluding 1. incorrect or unlikely bugs; 2. non-semantic bugs. Check the left bugs only break the intended functionality or lead to vulnerabilities. Answer if the bugs are related to single or a few tokens. \\
\textbf{Round 3 Properties}: \texttt{code\_line*} + \texttt{explanation*} + \texttt{category} \\
\textbf{Round 3 Prompt}:
Now, classify this bug into these categories using the code and previous contexts: Security Vulnerability, Logic Bug, Enhancement, Unexpected Behavior, Symbol Not Defined, Module Not Imported, Bad Smell, Not a Bug, or Others. If you assign others, explicitly specify its category.
\end{tcolorbox}
\caption{Prompt template 1/2FT/3Ca.}
\label{fig:template_12FM3Ca}
\end{figure}

\begin{figure}[h]
\begin{tcolorbox}[colback=gray!10!white, 
                  colframe=black, 
                  fonttitle=\small,
                  fontupper=\small,
                  fontlower=\small,
                  arc=2mm, 
                  boxsep=5pt, 
                  left=0pt, 
                  right=0pt, 
                  top=-2pt,
                  bottom=-2pt,
                  boxrule=0.5pt 
                  ]
\textbf{Round 1 Properties}: \texttt{code\_line} + \texttt{explanation} \\
\textbf{Round 1 Prompt}: \texttt{\{code\}}
Output exact lines of semantic bugs, concise explanations of the bugs, the fixed line of the bugs, and if the bugs can be fixed by merely changing a variable/method name. \\
Also, pay additional attention to these lines: \{suspicious\_lines\} \\
\textbf{Round 2 Properties}: \texttt{code\_line*} + \texttt{explanation*} + \texttt{fixed\_line} + \texttt{token\_level} + \texttt{category} \\
\textbf{Round 2 Prompt}:
Inspect these bugs, excluding 1. incorrect or unlikely bugs; 2. non-semantic bugs. Check the left bugs only break the intended functionality or lead to vulnerabilities. Finish the following tasks: 1. Answer if the bugs is related to single or a few tokens. 2. Using the code and previous contexts, classify the bugs into categories : Security Vulnerability, Logic Bug, Enhancement, Unexpected Behavior, Symbol Not Defined, Module Not Imported, Bad Smell, Not a Bug, or Others. Note that the snippet is from popular repositories and runs, so correct symbols not defined in the snippet might be defined at other places and you can regard it as Symbol not Defined. If you assign Others category, explicitly name the category.
\end{tcolorbox}
\caption{Prompt template 1/2FTCa w/ HL.}
\label{fig:template_12FMCaHL}
\end{figure}

\end{document}